\documentclass[11pt, a4paper]{article}

\usepackage{amsmath, cases, url}

\usepackage[dvips]{graphicx, color}

\newcommand{\pr}{p_\rightarrow}
\newcommand{\pl}{p_\leftarrow}
\newcommand{\pp}{p_0}

\newcommand{\classicP}{\tilde{P}} % probability for the classical problem

\begin{document}

\title{Hypergeometric solution to a gambler's ruin problem with a nonzero halting probability}
\author{Ken Yamamoto}
\date{}
\maketitle
{Department of Physics, Faculty of Science and Engineering,
	Chuo University, Kasuga, Bunkyo, Tokyo 112-8551, Japan}

{E-mail:} \url{yamamoto@phys.chuo-u.ac.jp}

\begin{abstract}
This paper treats of a kind of a gambler's ruin problem,
which seeks the probability that a random walker first hits the origin
at a certain time.
In addition to a usual random walk which hops either rightwards or leftwards,
the present paper introduces the `halt'
that the walker does not hop with a certain probability.
The solution to the problem can be obtained exactly
using a Gauss hypergeometric function.
The moment generating function of the duration is also calculated,
and a calculation technique of the moments is developed.
The author derives the long-time behavior of the ruin probability,
which exhibits power-law behavior
if the walker hops to the right and left with equal probability.
\end{abstract}



\section{Introduction}
% About the first-passage problem
The {\it gambler's ruin problem} is a classical subject of probability theory.
Consider a random walker on a one-dimensional lattice
hopping to the right with probability $\pr$ and left with $\pl(=1-\pr)$
in a single time step.
The problem seeks statistical properties of duration $T_x$,
the time at which the walker from $x$ first hits the origin.
(Of course, $T_x$ is a random variable.)
Not only a model of bankruptcy \cite{Scott} and gambling \cite{Maitra},
but queuing theory \cite{Keilson} and genetic algorithm \cite{Harik}
are concerned with the gambler's ruin problem.

A standard textbook of probability theory \cite{Feller}
gives detailed instructions on this problem.
The central quantity is
the probability $\classicP(x,t)$ that the walker at position $x$
first hits the origin after time $t$.
As the solution of the equation
$\classicP(x,t+1)=\pr \classicP(x+1,t)+\pl \classicP(x-1,t)$
with initial and boundary conditions
$\classicP(0,0)=1$ and $\classicP(x,0)=\classicP(0,t)=0$ ($x,t\ge1$),
\begin{equation}
\classicP(x,t)=
\begin{cases}
\displaystyle\frac{x}{t}\binom{t}{\frac{t+x}{2}}\pr^{(t-x)/2}\pl^{(t+x)/2}
	& \text{$t$ and $x$ are of same parity,}\\
0 & \text{$t$ and $x$ are of the opposite parity}.
\end{cases}
\label{eq:classic}
\end{equation}
The coefficient before $\pr^{(t-x)/2}\pl^{(t+x)/2}$ is
the number of different paths from $x$ hitting the origin first at time $t$,
and it is connected with the reflection principle of a random walk
\cite{Stanton}.
Obviously, $\classicP(x,t)=0$ holds when $t<x$,
because $\binom{t}{\frac{t+x}{2}}=0$.

% physics background, and `first passage problem'
In statistical physics,
a similar problem is called the {\it first passage problem} \cite{Redner}.
This is a more general problem than the ruin problem,
in that the state space of a random walker (or a diffusion particle)
is not necessarily a one-dimensional lattice
but higher-dimensional spaces and networks.
Many fields in statistical physics, including
reaction-rate theory \cite{Hanggi},
neuron dynamics \cite{Verechtchaguina},
and economic analysis \cite{Sazuka},
have been formulated and analyzed based on first-passage properties.

\begin{figure}[t!]
\centering
\includegraphics[clip]{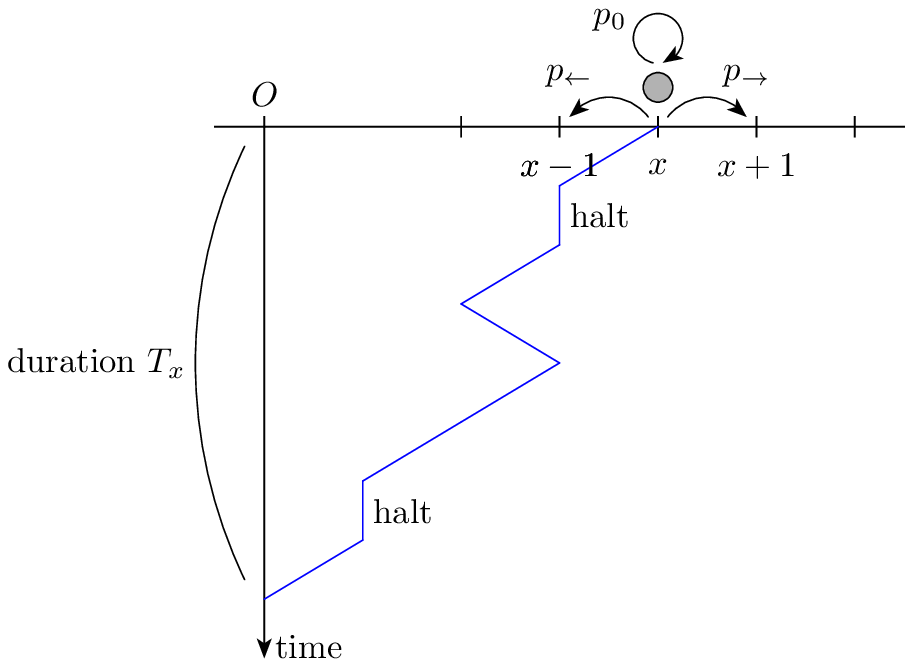}
\caption{
An illustration of the problem.
The random walker hops to the right with probability $\pr$,
and to the left with $\pl$;
hopping does not occur with probability $\pp$.
($\pr+\pl+\pp=1$.)
The problem focuses on statistical properties of the duration $T_x$,
which is the time the walker first hits the origin.
}
\label{fig1}
\end{figure}

The present paper analyzes an extended form of the classical ruin problem;
a random walker hops to the right with probability $\pr$,
to the left with $\pl$, and it does not hop with probability $\pp=1-\pr-\pl$.
Figure \ref{fig1} schematically shows the problem.
The only difference from the original ruin problem is that
the {\it halting} probability $\pp$ is introduced.
The difference seems very small,
but the results become quite distinct.
In fact, the solution \eqref{eq:classic} of non halting case ($\pp=0$)
is superseded by a formula involving a Gauss hypergeometric function
in halting case $\pp\ne0$.
Moreover, moment analysis and asymptotic (long-time) behavior
are developed.

\section{Hypergeometric solution} \label{sec:2}
As in the classical problem described in the previous section,
let $P(x,t)$ be the probability
that the walker starting from $x$ has duration $T_x=t$.
$P(x,t)$ satisfies the equation
$P(x,t+1)=\pr P(x+1,t)+\pl P(x-1,t)+\pp P(x,t)$
with initial and boundary conditions
$P(0,0)=1$ and $P(x,0)=P(0,t)=0$ ($x,t\ge1$).
However, it is hard to solve this equation directly,
and we take an another way
by employing a classical result \eqref{eq:classic} effectively.

To calculate $P(x,t)$,
we classify the walker's paths according to the number of hopping.
We count the paths consisting of $j$ hops and $t-j$ halts.
First, the different patterns of putting $t-j$ halts into $t$ steps
are given by $\binom{t-1}{t-j}$ in total,
where `$t-1$' (not $t$) comes from the fact
that a halting step never comes to the last $t$-th step.
Next, if we focus on only the hopping steps (and ignore the halting steps),
the paths are reduced to those of classical ruin problem;
the probability that the walker from $x$ hits the origin
after $j$ hops is given by $\classicP(x,j)$.
Thus the total occurring probability of a path
with $j$ hops and $t-j$ halts is
$$
\binom{t-1}{t-j}\pp^{t-j}\classicP(x,j).
$$
Summing up all $j$, we get
\begin{align}
P(x,t)
&= \sum_{j=x}^t \binom{t-1}{t-j}\pp^{t-j}\classicP(x,j) \nonumber\\
&= \sum_{\substack{j=x\\ \text{$j+x$ even}}}^t
   \pr^{(j-x)/2}\pl^{(j+x)/2}\pp^{t-j}x
   \frac{(t-1)!}{\left(\frac{j+x}{2}\right)!\left(\frac{j-x}{2}\right)!(t-j)!}.
\label{eq:P}
\end{align}
In the third equality, the range of summation is changed;
the terms corresponding to $j>t$ have no contribution
because $\frac{1}{(t-j)!}=0$.
In the last equality we change the summation variable as $j=x+2k$.
This result can be also obtained by solving directly
the equation $P(x,t+1)=\pr P(x+1,t)+\pl P(x-1,t)+\pp P(x,t)$
with conditions $P(0,0)=1$, and $P(x,0)=P(0,t)=0$ ($x\ge1$, $t\ge1$).

We further proceed with calculation;
we rewrite factorials using gamma functions,
which is a preparatory step toward a hypergeometric function.
Employing some formulas of the gamma function, one obtains
\begin{equation}
\frac{1}{(t-x-2k)!}
=\frac{1}{\Gamma(t+1-x)}
	\frac{\Gamma(\frac{x-t}{2}+\frac{1}{2}+k)\Gamma(\frac{x-t}{2}+k)}
		{\Gamma(\frac{x-t}{2}+\frac{1}{2})\Gamma(\frac{x-t}{2})}
 	2^{2k}.
\label{eq:gammaexpression}
\end{equation}
(See Appendix \ref{apdx:A} for the calculation in detail.)
The probability $P(x,t)$ is expressed as
\begin{align}
P(x,t)
&=\pl^x\pp^{t-x}x\frac{(t-1)!}{\Gamma(x+1)\Gamma(t-x+1)} \nonumber\\
&\qquad
	\frac{\Gamma(x+1)}
		{\Gamma(\frac{x-t}{2}+\frac{1}{2})\Gamma(\frac{x-t}{2})}
	\sum_{k=0}^\infty
	\frac{\Gamma(\frac{x-t}{2}+k+\frac{1}{2})\Gamma(\frac{x-t}{2}+k)}
		{\Gamma(x+k+1)}
	\frac{1}{k!}\left(\frac{4\pr\pl}{\pp^2}\right)^k\nonumber\\
&=\pl^{x}\pp^{t-x}\frac{(t-1)!}{(x-1)!(t-x)!}
	F\left(\frac{x-t}{2},\frac{x-t+1}{2}; x+1;\frac{4\pr\pl}{\pp^2}\right),
\label{eq:solution}
\end{align}
where $F(\alpha,\beta;\gamma;z)$ is the Gauss hypergeometric function.
This is an explicit form of the solution of our problem.
This solution cannot be deduced
from the non-halting solution \eqref{eq:classic}.
The hypergeometric function in Eq. \eqref{eq:solution}
is a {\it genuine} hypergeometric function,
in the sense that it cannot be expressed using simpler functions.
(It has been studied that
some hypergeometric functions have tractable expressions,
e.g., $F(1/2,1;3/2;-z^2)=z^{-1}\arctan z$. \cite{Abramowitz})

We comment here on the convergence of the sum in Eq. \eqref{eq:solution}---%
at first sight, it seems to diverge when $4\pr\pl/\pp^2>1$.
By using the Pochhammer symbol $(\alpha)_k:=\alpha(\alpha+1)\cdots(\alpha+k-1)$
instead of gamma functions,
$$
P(x,t)=\pl^x \pp^{t-x}\frac{(t-1)!}{(x-1)!(t-x)!}
\sum_{k=0}^\infty\frac{(\frac{x-t+1}{2})_k (\frac{x-t}{2})_k}{(x+1)_k}
	\frac{1}{k!}\left(\frac{4\pr\pl}{\pp^2}\right)^k.
$$
If $x>t$, $P(x,t)=0$ holds automatically because $1/(t-x)!=0$.
On the other hand, if $x\le t$,
either $(\frac{x-t}{2})_k$ or $(\frac{x-t+1}{2})_k$ becomes zero
for $k>\lfloor\frac{t-x}{2}\rfloor$.
(More precisely,
the former becomes zero when $x$ and $t$ are of same parity,
and the latter becomes zero otherwise.)
The sum consists of a finite number of terms in reality,
so one does not need to worry about the convergence.

\begin{figure}[tb!]\centering
\begin{minipage}{0.45\textwidth}\centering
\includegraphics[clip]{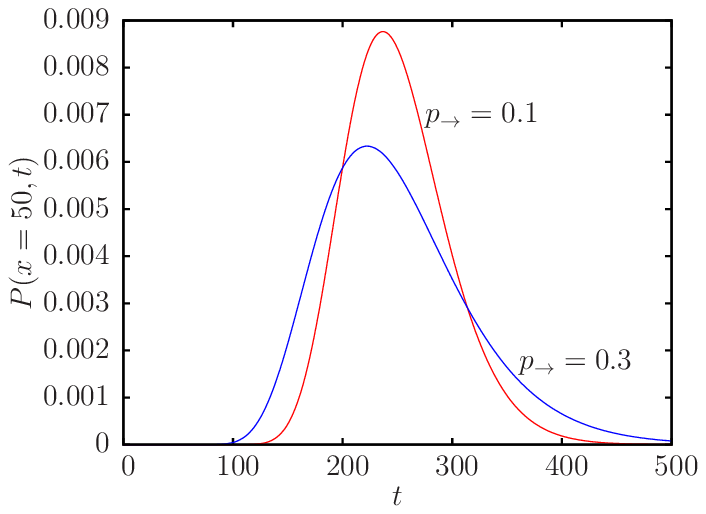}\\ (a)
\end{minipage}
\begin{minipage}{0.45\textwidth}\centering
\includegraphics[clip]{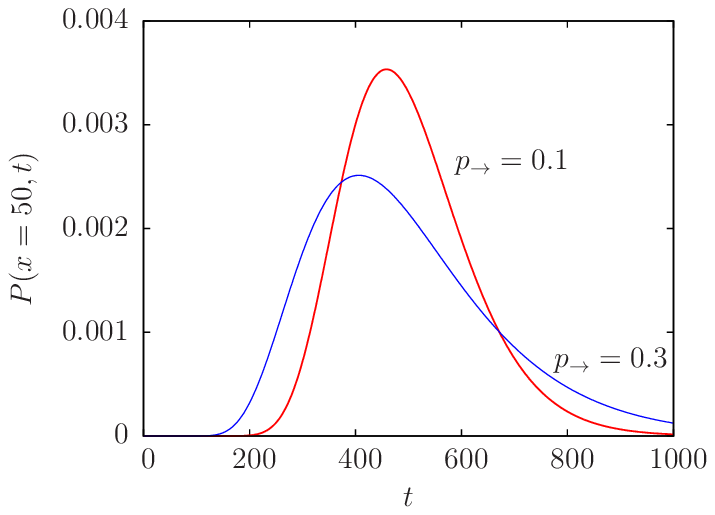}\\ (b)
\end{minipage}
\\[5mm]
\begin{minipage}{0.45\textwidth}\centering
\includegraphics[clip]{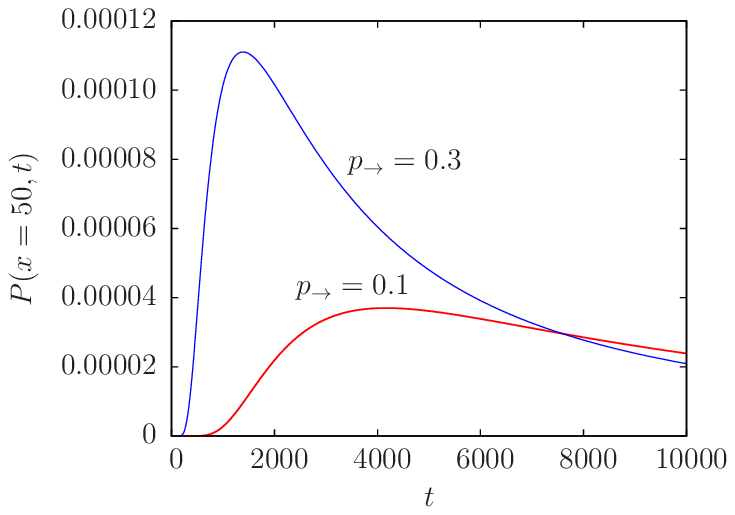}\\ (c)
\end{minipage}
\begin{minipage}{0.45\textwidth}\centering
\includegraphics[clip]{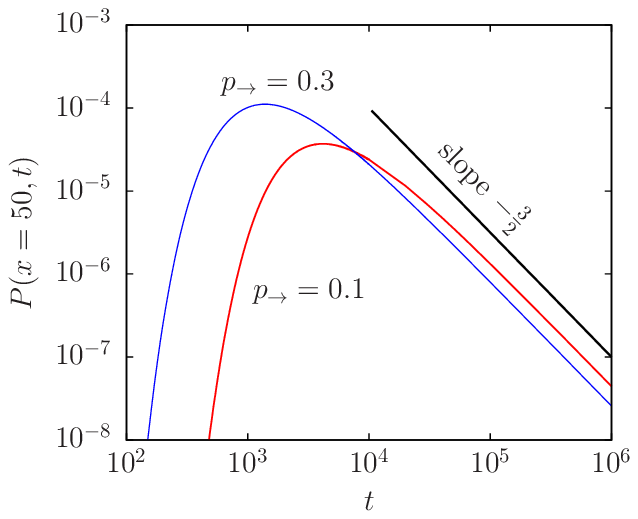}\\ (d)
\end{minipage}
\caption{
Numerical results of $P(x,t)$ as a function of $t$ ($x=50$).
The graphs are separated according to the value of $\varDelta p=\pl-\pr$:
(a) $\varDelta p=0.2$, (b) $\varDelta p=0.1$, and (c) $\varDelta p=0$.
The two curves in each panel represent $\pr=0.1$ and $0.3$.
Long-time behavior of $P(x,t)$ in $\varDelta p=0$
is shown in (d) on logarithmic scales,
which suggests a power law $P(x,t)\propto t^{-3/2}$.
}
\label{fig2}
\end{figure}

The exact solution \eqref{eq:solution} is difficult to understand intuitively.
We show numerical evaluation of $P(x,t)$ in Fig. \ref{fig2}.
The initial position of the walker is fixed as $x=50$.
We separate graphs according to the parameter $\varDelta p:=\pl-\pr$;
the panels (a), (b), and (c) respectively correspond to
$\varDelta p=0.2$, $0.1$, and $0$.
($\varDelta p$ is a key parameter for the average duration,
as discussed in the following section.)
$P(x,t)$ is a unimodal function of $t$ for each parameter value.
Power-law behavior $P(x,t)\propto t^{-3/2}$ is suggested in large $t$
when $\varDelta p=0$ (see (d)),
which is further discussed in Sec. \ref{sec:asymptotic}.

\section{Analysis of moment}\label{sec:3}
In order to see properties of the random variable $T_x$,
moment analysis is developed in this section.

First, we calculate the moment generating function of $T_x$, defined as
\begin{equation}
M_x(s)=\sum_{t=0}^\infty e^{st}P(x,t).
\label{eq:definitionM}
\end{equation}
Calculation process is summarized in Appendix \ref{apdx:B}, and the result is
\begin{equation}
M_x(s)=\left(
		\frac{e^{-s}-\pp}{2\pr}
		-\sqrt{\left(\frac{e^{-s}-\pp}{2\pr}\right)^2-\frac{\pl}{\pr}}
	\right)^x.
\label{eq:momentgenerating}
\end{equation}
$M_x(s)$ contains no special functions,
so this is simpler than $P(x,t)$ of Eq. \eqref{eq:solution}.
Remarkably,
$M_x(s)$ is described by elementary functions, though
its definition \eqref{eq:definitionM} is an infinite sum
of which each term contains a genuine hypergeometric function
Remarkably,
$M_x(s)$ contains no special functions, though
the infinite summation in Eq. \eqref{eq:definitionM}
involves a genuine hypergeometric function via $P(x,t)$.

Repeatedly differentiating $M_x(s)$ and putting $s=0$,
we have the first and second moments of the duration $T_x$ as
\begin{align*}
\langle T_x\rangle &=\frac{x}{\pl-\pr},\\
\langle T_x^2\rangle &=\frac{x^2}{(\pl-\pr)^2}
	+\left\{
		\frac{\pl+\pr}{(\pl-\pr)^3}-\frac{1}{\pl-\pr}
	\right\}x.
\end{align*}
The average and the other moments remain finite when $\pl>\pr$,
and they diverge to infinity as $\pl-\pr\searrow0$.
The variance of $T_x$ is given by
$$
V(T_x):=\langle T_x^2\rangle-\langle T_x\rangle^2
=\left\{\frac{\pl+\pr}{(\pl-\pr)^3}-\frac{1}{\pl-\pr}\right\}x.
$$

We can give a plain explanation for the average $\langle T_x\rangle$.
The walker moves to the left by a length $\varDelta p=\pl-\pr$
on average in a single hop.
In other words, $\varDelta p$
is a mean velocity of the walker toward the origin.
Hence, it takes $x/\varDelta p$ time steps on average to reach the origin.

The moment generating function \eqref{eq:momentgenerating}
looks too complicated to calculate higher moments by differentiation.
Alternatively,
we provide another calculation method for moments of the duration $T_x$.

We start from the difference equation for $P(x,t)$:
$$
P(x,t+1)=\pr P(x+1,t)+\pl P(x-1,t)+\pp P(x,t).
$$
Multiply $t$ and take summation for $t$ to get
\begin{align*}
\sum_{t=0}^\infty tP(x,t+1)
&=\sum_{t=0}^\infty \{t\pr P(x+1,t)+t\pl P(x-1,t)+t\pp P(x,t)\}\\
&=\pr\langle T_{x+1}\rangle+\pl\langle T_{x-1}\rangle+\pp\langle T_x\rangle.
\end{align*}
The left-hand side can be expressed as
$$
\sum_{t=0}^\infty tP(x,t+1)
=\sum_{t=0}^\infty (t+1)P(x,t+1)-\sum_{t=0}^\infty P(x,t+1)
=\langle T_x\rangle-1.
$$
The last equality uses $\sum_{t=0}^\infty P(x,t+1)=1$,
which means that a random walker surely hits the origin.
(This normalization breaks if $\pl<\pr$---%
see the discussion of total probability of ruin in Sec. \ref{sec:discussion}
in more detail.)
We have a difference equation
$$
\pr\langle T_{x+1}\rangle-(\pr+\pl)\langle T_x\rangle+\pl\langle T_{x-1}\rangle
=-1.
$$
A particular solution is given by $\langle T_x\rangle=x/(\pl-\pr)$,
and the characteristic equation $\pr\lambda^2-(\pr+\pl)\lambda+\pl=0$
admits the two roots $1$ and $\pl/\pr$.
Thus, the general solution is
$$
\langle T_x\rangle = \frac{x}{\pl-\pr}+A+B\left(\frac{\pl}{\pr}\right)^x,
$$
which has two constants $A$ and $B$.
In a limiting case $\pr\to0$,
the problem is still well-defined, and $\langle T_x\rangle$ should be finite,
so $B=0$.
Also, it is obvious that $\langle T_0\rangle=0$, so $A=0$.
The appropriate solution is therefore
$$
\langle T_x\rangle=\frac{x}{\pl-\pr}.
$$

Next we calculate the second moment.
Multiplying $t^2$ and taking summation, we derive a difference equation
$$
\langle T_x^2\rangle-2\langle T_x\rangle+1
=\pr\langle T_{x+1}^2\rangle+\pl\langle T_{x-1}^2\rangle
	+\pp\langle T_x^2\rangle,
$$
where we use $t^2=(t+1)^2-2(t+1)+1$ for the calculation on the left-hand side.
We assume a particular solution in the form
$\langle T_x^2\rangle=C_2 x^2+C_1 x$,
and determine two constants as
$$
C_1 =\frac{\pl+\pr}{(\pl-\pr)^3}-\frac{1}{\pl-\pr},\quad
C_2 = \frac{1}{(\pl-\pr)^2}.
$$
The second moment is
$$
\langle T_x^2\rangle
=\frac{x^2}{(\pl-\pr)^2}
+\left\{\frac{\pl+\pr}{(\pl-\pr)^3}-\frac{1}{\pl-\pr}\right\}x
=\frac{x^2}{(\pl-\pr)^2}
+\frac{\pp(1-\pp)+4\pr\pl}{(\pl-\pr)^3}x.
$$
The homogeneous-solution part vanishes for the same reason as the average.

In short, this `bottom-up' method calculates higher moments inductively,
and generally $\langle T_x^{k}\rangle$ becomes a polynomial of degree $k$.
For instance, 
the third moment is calculated as
\begin{align*}
\langle T_x^3\rangle
&=\frac{x^3}{(\pl-\pr)^3}
+\left\{\frac{\pl+\pr}{(\pl-\pr)^3}-\frac{1}{\pl-\pr}\right\}3x^2\\
&\quad+\left\{\frac{2(\pl+\pr)^2+4\pl\pr}{(\pl-\pr)^5}
	-\frac{3(\pl+\pr)}{(\pl-\pr)^3}+\frac{1}{\pl-\pr}\right\}x.
\end{align*}

\section{Asymptotic behavior}\label{sec:asymptotic}
We look at long-time behavior of $P(x,t)$ in this section.
In particular, we confirm the power-law behavior $P(x,t)\propto t^{-3/2}$
mentioned in Sec. \ref{sec:2} when $\pr=\pl$
(see Fig. \ref{fig2} (d) for reference).

A summation and factorial, as well as a hypergeometric function,
are not suitable to study limiting behavior,
so we first reexpress the probability $P(x,t)$ into a tractable form.
(It may also be possible to estimate factorials directly
by Stirling's formula.)
According to Ref. \cite{Feller},
the solution \eqref{eq:classic} of the classic ruin problem
has another expression
\begin{equation}
\classicP(x,t)=2^t\pr^{(t-x)/2}\pl^{(t+x)/2}
\int_0^1\cos^{t-1}\pi\phi\cdot\sin\pi\phi\cdot\sin\pi x\phi d\phi.
\label{eq:integralclassic}
\end{equation}
This form is convenient in that
one does not need to care about the parities of $x$ and $t$.
In fact, the integral holds the information of parities:
\begin{equation}
\int_0^1\cos^{t-1}\pi\phi\cdot\sin\pi\phi\cdot\sin\pi x\phi d\phi
=
\begin{cases}
\displaystyle\frac{x}{2^t}
	\frac{(t-1)!}{(\frac{t+x}{2})!(\frac{t-x}{2})!}
	& \text{$t$ and $x$ are of same parity,}\\
0 & \text{otherwise.}
\end{cases}
\label{eq:integralform}
\end{equation}
In Appendix \ref{apdx:C} we prove this formula.
Applying Eq. \eqref{eq:integralclassic} to Eq. \eqref{eq:P},
we have the integral form of the probability $P(x,t)$:
\begin{align*}
P(x,t)
&=\sum_{j=1}^t \binom{t-1}{t-j}\pp^{t-j}\classicP(x,j)\\
&= 2\pr^{(1-x)/2}\pl^{(1+x)/2}
	\int_0^1
		\left\{
			\sum_{j=1}^{t}\binom{t-1}{t-j}\pp^{t-j}
			(2\sqrt{\pr\pl}\cos\pi\phi)^{j-1}
		\right\}
	\sin\pi\phi\cdot\sin\pi x\phi d\phi\\
&= 2\pr^{(1-x)/2}\pl^{(1+x)/2}
	\int_0^1
	(\pp+2\sqrt{\pr\pl}\cos\pi\phi)^{t-1}
	\sin\pi\phi\cdot\sin\pi x\phi d\phi\\
&= 2\pr^{(1-x)/2}\pl^{(1+x)/2}
	(\pp+2\sqrt{\pr\pl})^{t-1}
	\int_0^1
	\left(
		\frac{\pp+2\sqrt{\pr\pl}\cos\pi\phi}{\pp+2\sqrt{\pr\pl}}
	\right)^{t-1}
	\sin\pi\phi\cdot\sin\pi x\phi d\phi.
\end{align*}

Here we estimate the integral
\begin{align*}
I&:=
\int_0^1
	\left(
		\frac{\pp+2\sqrt{\pr\pl}\cos\pi\phi}{\pp+2\sqrt{\pr\pl}}
	\right)^{t-1}
	\sin\pi\phi\cdot\sin\pi x\phi d\phi\\
&=\int_0^{\pi t}
	\left(
		\frac{\pp+2\sqrt{\pr\pl}\cos\frac{\nu}{t}}{\pp+2\sqrt{\pr\pl}}
	\right)^{t-1}
	\sin\frac{\nu}{t}\cdot\sin\frac{x\nu}{t}\cdot\frac{1}{\pi t} d\nu,
\end{align*}
where the integral variable is changed as $\nu=\pi t\phi$
in the second equality.
For large $t$,
$$
\left(
	\frac{\pp+2\sqrt{\pr\pl}\cos\frac{\nu}{t}}{\pp+2\sqrt{\pr\pl}}
\right)^{t-1}
\simeq \left(
	1-\frac{\sqrt{\pr\pl}}{\pp+2\sqrt{\pr\pl}}\frac{\nu^2}{t^2}
\right)^{t-1}
\simeq \exp\left(
	-\frac{\sqrt{\pr\pl}}{\pp+2\sqrt{\pr\pl}}\frac{\nu^2}{t}
\right).
$$
Hence the integrand is a rapidly decreasing function of $\nu$,
and we can extend the upper limit of the integration to infinity.
Together with the approximation $\sin(\nu/t)\simeq\nu/t$,
we can carry out the integration as
$$
I\simeq \frac{x}{\pi t^3}
\int_0^\infty 
\exp\left(
	-\frac{\sqrt{\pr\pl}}{\pp+2\sqrt{\pr\pl}}\frac{\nu^2}{t}
\right)
\nu^2 d\nu
= \frac{x}{4\sqrt{\pi}}
\left(\frac{\pp+2\sqrt{\pr\pl}}{\sqrt{\pr\pl}}\right)^{3/2}
t^{-3/2}.
$$

Therefore, the asymptotic form is expressed as
$$
P(x,t)\simeq \frac{x}{2\sqrt{\pi}}
\pr^{(1-x)/2}\pl^{(1+x)/2}(\pp+2\sqrt{\pr\pl})^{t-1}
\left(\frac{\pp+2\sqrt{\pr\pl}}{\sqrt{\pr\pl}}\right)^{3/2}
t^{-3/2}.
$$
By the arithmetic mean-geometric mean inequality, we note
$$
\pp+2\sqrt{\pr\pl}\le\pp+\pr+\pl=1,
$$
with equality holding if and only if $\pr=\pl$
(or $\varDelta p=0$ by the notation in Sec. \ref{sec:2}).
Thus, the long-time behavior of $P(x,t)$ is completely different
according to whether $\varDelta p$ equals zero or not.
\begin{itemize}
\item If $\varDelta p\ne0$ (i.e., $\pr\ne\pl$),
$\pp+2\sqrt{\pr\pl}<1$, and hence $P(x,t)$ asymptotically exhibits
exponential decay due to $(\pp+2\sqrt{\pr\pl})^t$.
\item If $\varDelta p=0$ (i.e., $\pr=\pl=:p$),
a power law with exponent $-3/2$
$$
P(x,t)\simeq \frac{x}{2\sqrt{\pi p}}t^{-3/2}
$$
is concluded (recall Fig. \ref{fig2} (d)).
\end{itemize}

\section{Concluding remarks and discussions} \label{sec:discussion}
% summary
In the present paper,
we have solved a gambler's ruin problem where a random walker hops
to the right with probability $\pr$, left with $\pl$,
and does not hop with $\pp$.
We have calculated exactly the probability $P(x,t)$
that the walker starting from position $x$ has duration $T_x=t$.
The average and higher moments of $T_x$ are calculated in two ways:
by the moment generating function,
and by the bottom-up calculation.
The asymptotic form of $P(x,t)$ is derived,
and a power law $P(x,t)\sim t^{-3/2}$ is obtained when $\pr=\pl$.

We make a brief discussion on the continuum limit,
which is an appropriate scaling limit where the step width and time interval
of the walker tend to zero.
Let the step width be $\delta$ and time interval be $\epsilon$,
and position and time are scaled as $\xi=\delta x$ and $\tau=\epsilon t$.
In the continuum limit we take
$\delta,\epsilon\to0$, together with
$$
(\pr-\pl)\frac{\delta}{\epsilon}\to v,\quad
\frac{\delta^2}{2\epsilon}\to D
$$
are both kept finite.
$v$ is the mean displacement per time unit,
and $D$ is the diffusioin coefficient
corresponding to a non-halting (or formally called a {\it simple}) random walk.
The probability density function
$$
\rho(\xi,\tau)
=\lim\frac{1}{\epsilon}P\left(\frac{x}{\delta},\frac{t}{\epsilon}\right)
$$
is suitable in the continuum limit,
rather than the probability distribution $P(x,t)$.
(`$\lim$' represents the continuum limit.)
Carrying out calculation similar to that in Ref. \cite{Feller},
$$
\rho(\xi,\tau)=\frac{\xi}{\sqrt{4\pi(1-\pp)D\tau^3}}
	\exp\left(-\frac{(\xi+v\tau)^2}{4(1-\pp)D\tau}\right)
$$
is obtained.
The probability density $\rho$ is called
the inverse Gaussian distribution \cite{Tweedie}.
In the continuum limit, the halting effect $\pp$ is reflected
only upon the diffusion coefficient as $(1-\pp)D$.
Comparing $P(x,t)$ of a discrete problem in Eq. \eqref{eq:solution}
and above $\rho(\xi,\tau)$ of a continuum limit,
we conclude that the discrete random walk is far more difficult
than the continuous diffusion,
and that the results about the discrete random walk in this paper
cannot be attained from continuous diffuision problem.

Combining Eqs. \eqref{eq:solution}, \eqref{eq:definitionM},
and \eqref{eq:momentgenerating}, we get
$$
\sum_{t=0}^\infty e^{st}\pl^x\pp^{t-x}\binom{t-1}{x-1}
	F\left(\frac{x-t}{2},\frac{x-t+1}{2}; x+1; \frac{4\pr\pl}{\pp^2}\right)
=\left(
	\frac{e^{-s}-\pp}{2\pr}
	-\sqrt{\left(\frac{e^{-s}-\pp}{2\pr}\right)^2-\frac{\pl}{\pr}}
\right)^x.
$$
We have not used the condition $\pr+\pl+\pp=1$
in the derivation of Eqs. \eqref{eq:solution} and \eqref{eq:momentgenerating},
so $\pr$, $\pl$, and $\pp$ can take any value independently.
Set $\pr=z/4$ and $\pl=\pp=1$,
$$
\sum_{t=0}^\infty e^{st}\binom{t-1}{x-1}
	F\left(\frac{x-t}{2},\frac{x-t+1}{2}; x+1; z\right)
=\left(
	\frac{2(e^{-s}-1)}{z}
	-\sqrt{\left(\frac{2(e^{-s}-1)}{z}\right)^2-\frac{4}{z}}
\right)^x.
$$
Furthermore, set $s=-\ln 2$ (i.e., $e^{-s}=2$),
$$
\sum_{t=0}^\infty \frac{1}{2^t}\binom{t-1}{x-1}
	F\left(\frac{x-t}{2},\frac{x-t+1}{2}; x+1; z\right)
=\left(\frac{2-2\sqrt{1-z}}{z}\right)^x.
$$
These formulas are nontrivial,
but the author cannot tell whether they are useful in practice.
We stress that
an infinite series
of which each term involves a genuine hypergeometric function
is hardly known.

We comment on the total probability of ruin.
For the discrete problem,
putting $s=0$ the moment generating function $M_x(s)$,
\begin{align*}
\sum_{t=0}^\infty P(x,t)&=M_x(0) \nonumber\\
&=\left(
	\frac{1-\pp}{2\pr}
	-\sqrt{\left(\frac{1-\pp}{2\pr}\right)^2-\frac{\pl}{\pr}}
\right)^x \nonumber\\
&=\left(
	\frac{\pr+\pl}{2\pr}
	-\frac{\sqrt{(\pl-\pr)^2}}{2\pr}
\right)^x \nonumber\\
&=\begin{cases}
	1 & \text{$\pl\ge\pr$},\\
	\left(\displaystyle\frac{\pl}{\pr}\right)^x & \text{$\pl<\pr$}.
\end{cases}
\end{align*}
If $\pl>\pr$, the gambler surely comes to ruin, but
If $\pl<\pr$, the gambler can manage to avoid running out of the bankroll
with nonzero probability.

The equation $P(x,t+1)=\pr P(x+1,t)+\pl P(x-1,t)+\pp P(x,t)$
describes a traffic jam,
where $P(x,t)$ stands for the probability distribution
that the jam consists of $x$ cars at time $t$,
and $\pr$ and $\pl$ are involved in the rates
at which cars enter and leave the jam.
The scaling behavior with exponent $-3/2$ has been observed
in the lifetime distribution of jams \cite{Nagel}.
We think that results obtained in this paper can become theoretical bases;
in particular,
our discrete problem is comparable to a microscopic description of traffic,
where discreteness is not negligible.

\section*{Acknowledgment}
The author is grateful to Dr. Yoshihiro Yamazaki and Dr. Jun-ichi Wakita
for their beneficial comments.

%%%%%%%%%%%%%%%%%%%%%%%%%%%%%%%%%%%%%%%%%%%%%%%%%%
% Regular sections end
% Appendices start
\appendix
%%%%%%%%%%%%%%%%%%%%%%%%%%%%%%%%%%%%%%%%%%%%%%%%%%

\section{Calculation of Eq. \eqref{eq:gammaexpression}} \label{apdx:A}
Here we follow the calculation of Eq. \eqref{eq:gammaexpression} in detail.
Employing the duplication formula
$\Gamma(2z)=\pi^{-1/2}2^{2z-1}\Gamma(z)\Gamma(z+\frac{1}{2})$,
we rewrite
\begin{equation}
\frac{1}{(t-x-2k)!}=\frac{1}{\Gamma(t-x-2k+1)}
=\frac{\sqrt{\pi}}{2^{t-x-2k}\Gamma(\frac{t-x}{2}-k+\frac{1}{2})\Gamma(\frac{t-x}{2}-k+1)}.
\label{eq:A1}
\end{equation}
By Euler's reflection formula $\Gamma(z)\Gamma(1-z)=\pi/\sin\pi z$,
\begin{align}
\Gamma\left(\frac{t-x}{2}-k+\frac{1}{2}\right)
&=\frac{1}{\Gamma(k-\frac{t-x}{2}+\frac{1}{2})}
	\frac{\pi}{\sin\pi(\frac{t-x}{2}-k+\frac{1}{2})} \nonumber\\
&=\frac{1}{\Gamma(k-\frac{t-x}{2}+\frac{1}{2})}
	\frac{(-1)^k\pi}{\sin\pi(\frac{t-x}{2}+\frac{1}{2})} \nonumber\\
&=\frac{(-1)^k}{\Gamma(k-\frac{t-x}{2}+\frac{1}{2})}
	\Gamma\left(\frac{t-x}{2}+\frac{1}{2}\right)
	\Gamma\left(\frac{1}{2}-\frac{t-x}{2}\right).
\label{eq:A2}
\end{align}
Note that
$\sin\pi(\frac{t-x}{2}-k+\frac{1}{2})=(-1)^k\sin\pi(\frac{t-x}{2}+\frac{1}{2})$ because $k$ is an integer.
Similarly,
\begin{equation}
\Gamma\left(\frac{t-x}{2}-k+1\right)
=\frac{(-1)^k}{\Gamma(k+\frac{x-t}{2})}
\Gamma\left(\frac{t-x}{2}+1\right)\Gamma\left(\frac{x-t}{2}\right)
\label{eq:A3}
\end{equation}
is obtained.
Substituting Eqs. \eqref{eq:A2} and \eqref{eq:A3} into Eq. \eqref{eq:A1},
then using the duplication formula again as
$$
\Gamma\left(\frac{t-x}{2}+\frac{1}{2}\right)\Gamma\left(\frac{t-x}{2}+1\right)
=\frac{\sqrt{\pi}}{2^{t-x}}\Gamma(t-x+1),
$$
we eventually come to the result
$$
\frac{1}{(t-x-2k)!}
=\frac{1}{\Gamma(t+1-x)}
	\frac{\Gamma(\frac{x-t}{2}+\frac{1}{2}+k)\Gamma(\frac{x-t}{2}+k)}
		{\Gamma(\frac{x-t}{2}+\frac{1}{2})\Gamma(\frac{x-t}{2})}
 	2^{2k}.
$$

\section{Calculation of the moment generating function \eqref{eq:momentgenerating}} \label{apdx:B}
We follow here the calculation of Eq. \eqref{eq:momentgenerating}.
Substituting Eq. \eqref{eq:P} into Eq. \eqref{eq:definitionM},
\begin{align}
M_x(s)&=\sum_{t=0}^\infty e^{st}
	\sum_{k=0}^\infty \pr^k\pl^{x+k}\pp^{t-x-2k}x
	\frac{(t-1)!}{(x+k)!k!(t-x-2k)!} \nonumber\\
&=\sum_{k=0}^\infty \pr^k\pl^{x+k}\pp^{-x-2k}x\frac{1}{(x+k)!k!}
	\sum_{t=0}^\infty\frac{(t-1)!}{(t-x-2k)!}(\pp e^s)^t \nonumber\\
&=\sum_{k=0}^\infty \pr^k\pl^{x+k}\pp^{-x-2k}x
	\frac{1}{(x+k)!k!}(\pp e^s)^{x+2k}
	\sum_{\tau=0}^\infty\frac{(\tau+x+2k-1)!}{\tau!}(\pp e^s)^\tau,
\label{eq:moment}
\end{align}
where a summation variable is changed as $\tau=t-x-2k$ in the last equality.
Applying the negative binomial expansion
$$
(1-y)^{-m}=\sum_{\tau=0}^\infty \frac{(\tau+m-1)!}{\tau!(m-1)!}y^\tau
$$
with $y=\pp e^s$ and $m=x+2k$, one obtains
$$
\sum_{\tau=0}^\infty\frac{(\tau+x+2k-1)!}{\tau!}(\pp e^s)^\tau
=(x+2k-1)!(1-\pp e^s)^{-x-2k},
$$
and simplifies Eq. \eqref{eq:moment} into
$$
M_x(s)=\left(\frac{\pl}{e^{-s}-\pp}\right)^x x
\sum_{k=0}^\infty \frac{(x+2k-1)!}{(x+k)!k!}
\left(\frac{\sqrt{\pr\pl}}{e^{-s}-\pp}\right)^{2k}.
$$
We can transform as follows using the duplication formula
$\Gamma(2z)=\pi^{-1/2}2^{2z-1}\Gamma(z)\Gamma(z+\frac{1}{2})$,
\begin{align*}
(x+2k-1)!&=\Gamma(x+2k)\\
&=\frac{2^{x-1+2k}}{\sqrt{\pi}}
	\Gamma\left(\frac{x}{2}+k\right)\Gamma\left(\frac{x+1}{2}+k\right)\\
&=\frac{2^{x-1}}{\sqrt{\pi}}
	\Gamma\left(\frac{x}{2}\right)\Gamma\left(\frac{x+1}{2}\right)
	\frac{\Gamma\left(\frac{x}{2}+k\right)\Gamma\left(\frac{x+1}{2}+k\right)}
		{\Gamma\left(\frac{x}{2}\right)\Gamma\left(\frac{x+1}{2}\right)}
	2^{2k}\\
&=\Gamma(x)
	\frac{\Gamma\left(\frac{x}{2}+k\right)\Gamma\left(\frac{x+1}{2}+k\right)}
		{\Gamma\left(\frac{x}{2}\right)\Gamma\left(\frac{x+1}{2}\right)}
	2^{2k}.
\end{align*}
The moment generating function can be expressed by
a Gauss hypergeometric function
\begin{align*}
M_x(s)
&=\left(\frac{\pl}{e^{-s}-\pp}\right)^x
	\frac{x\Gamma(x)}{\Gamma(x+1)}
	\frac{\Gamma(x+1)}{\Gamma(\frac{x}{2})\Gamma(\frac{x+1}{2})}
	\sum_{k=0}^\infty
	\frac{\Gamma(\frac{x}{2}+k)\Gamma(\frac{x+1}{2}+k)}
		{\Gamma(x+1+k)}\frac{1}{k!}
	\left(
		\frac{2\sqrt{\pr\pl}}{e^{-s}-\pp}
	\right)^{2k}\\
&=\left(\frac{\pl}{e^{-s}-\pp}\right)^x
F\left(\frac{x}{2},\frac{x+1}{2}; x+1;\frac{4\pr\pl}{(e^{-s}-\pp)^2}\right).
\end{align*}
It can be expressed by an elementary function
(see Ref. \cite{Abramowitz}):
$$
F\left(\frac{x}{2},\frac{x+1}{2}; x+1;z\right)
=\left(\frac{2}{1+\sqrt{1-z}}\right)^x
=\left(\frac{2-2\sqrt{1-z}}{z}\right)^x.
$$
This leads to the conclusion.

\section{Proof of Eq. \eqref{eq:integralform}} \label{apdx:C}
We calculate the integral to prove Eq. \eqref{eq:integralform}.
We first break up the product of circular functions into the sum.
$$
\cos^{t-1}\pi\phi
=\left(\frac{e^{i\pi\phi}+e^{-i\pi\phi}}{2}\right)^{t-1}
=\frac{1}{2^{t-1}}
	\sum_{k=0}^{t-1}\binom{t-1}{k}e^{i(2k-t+1)\pi\phi},
$$
whose real part is
$$
\cos^{t-1}\pi\phi
=\frac{1}{2^{t-1}}\sum_{k=0}^{t-1}\binom{t-1}{k}\cos(2k-t+1)\pi\phi.
$$
This is just a Fourier series expansion of $\cos^{t-1}\theta$,
and is associated with the Chebyshev polynomial
$T_n(\cos \theta):=\cos n\theta$ \cite{Gil}.
Moreover, by formulas in trigonometry,
\begin{align*}
&\cos(2k-t+1)\pi\phi\cdot\sin\pi\phi\cdot\sin\pi x\phi\\
&=\frac{1}{4}\left\{
	\cos(2k-t+x)\pi\phi+\cos(2k+2-t-x)\pi\phi
	-\cos(2k+2-t+x)\pi\phi-\cos(2k-t-x)\pi\phi
\right\}.
\end{align*}
Thus,
\begin{align*}
&\int_0^1 \cos^{t-1}\pi\phi\cdot\sin\pi\phi\cdot\sin\pi x\phi d\phi\\
&=\frac{1}{2^{t+1}}\sum_{k=0}^{t-1}\binom{t-1}{k}\int_0^1
	\{\cos(2k-t+x)\pi\phi+\cos(2k+2-t-x)\pi\phi
	-\cos(2k+2-t+x)\pi\phi-\cos(2k-t-x)\pi\phi\}d\phi.
\end{align*}
Since $t$, $x$, and $k$ are integers, the first term in the integral is
$$
\int_0^1 \cos(2k-t+x)\pi\phi d\phi=
\begin{cases}
1 & 2k-t+x=0, \\
0 & \text{otherwise},
\end{cases}
$$
which behaves like the Kronecker delta
$\delta(k,\frac{t-x}{2})$---%
a subscript fraction may be unreadable,
so we use $\delta(i,j)$ instead of the standard symbol $\delta_{i,j}$.
The following terms in the integral are also substituted by
$\delta(k,\frac{t+x-2}{2})$,
$\delta(k,\frac{t-x-2}{2})$, and
$\delta(k,\frac{t+x}{2})$, respectively.

If $t$ and $x$ have opposite parity
(i.e., $t+x$ and $t-x$ are odd),
each Kronecker delta vanishes for all integer $k$.
Otherwise, if $t$ and $x$ have same parity,
each Kronecker delta becomes nonzero at some $k$.
By picking out such $k$,
\begin{align*}
\int_0^1 \cos^{t-1}\pi\phi\cdot\sin\pi\phi\cdot\sin\pi x\phi d\phi
&=\frac{1}{2^{t+1}}
\left\{
	\binom{t-1}{\frac{t-x}{2}}
	+\binom{t-1}{\frac{t+x-2}{2}}
	-\binom{t-1}{\frac{t-x-2}{2}}
	-\binom{t-1}{\frac{t+x}{2}}
\right\}\\
&=\frac{x}{2^t}\frac{(t-1)!}{(\frac{t+x}{2})!(\frac{t-x}{2})!}.
\end{align*}
Therefore, Eq. \eqref{eq:integralform} is derived.

%%%%%%%%%%%%%%%%%%%%%%%%%%%%%%%%%%%%%%%%%%%%%%%%%%
% References
%%%%%%%%%%%%%%%%%%%%%%%%%%%%%%%%%%%%%%%%%%%%%%%%%%

\end{document}